\documentclass{elsart4}

\usepackage{graphics,color}
\usepackage{graphicx}
\usepackage{epsfig}
\usepackage{amssymb}
\usepackage{latexsym}
\usepackage{bbm}
\usepackage{amstext}
\usepackage{amsxtra}

\journal{Physica B}

\usepackage{bm}

\newcommand{\bone}{\mathbbm{1}}

\newcommand{\K}{{\boldsymbol K}}
\renewcommand{\k}{{\boldsymbol k}}
\newcommand{\G}{{\boldsymbol G}}
\newcommand{\M}{{\boldsymbol M}}

\newcommand{\be}{\begin{equation}}
\newcommand{\ee}{\end{equation}}

\renewcommand{\a}{{\boldsymbol a}}

\newcommand{\D}{{\boldsymbol D}}

\newcommand{\q}{{\boldsymbol q}}

\newcommand{\R}{{\boldsymbol R}}

\newcommand{\ep}{\epsilon}

\begin{document}

\begin{frontmatter}


 \title{Manipulation of Dirac points in graphene-like crystals %\thanksref{label1}
 }
 %\thanks[label1]{}
 \author{R. de Gail, J.-N. Fuchs, M.-O. Goerbig, F. Pi\'echon and G. Montambaux\corauthref{cor1}%\thanksref{label2}
 }
 \address{Laboratoire de Physique des Solides, CNRS UMR 8502, Universit{\'e} Paris-Sud, 91405 Orsay, France %
% \thanksref{label3}
 }
 %\thanks[label3]{}
\corauth[cor1]{ e-mail: montambaux@lps.u-psud.fr}

%\title{}


%\author{}

%\address{}

%Version \today

\begin{abstract}
We review different scenarios for the motion and merging of Dirac points in two dimensional crystals. These different types of merging can be classified according to a   winding number (a topological Berry phase) attached to each Dirac point. For each scenario, we calculate the Landau level spectrum and show that it can be quantitatively described  by a semiclassical quantization rule for the  constant energy areas. This quantization depends on how many Dirac points are enclosed by these areas. We also emphasize that different scenarios are characterized by different numbers of topologically protected zero energy Landau levels.
\end{abstract}

\end{frontmatter}

% main text
\section{Introduction}

The spectacular properties of graphene \cite{Revs} have led to a tremendous interest for the study of physical properties related to an electronic spectrum exhibiting Dirac points with a massless spectrum. If, in graphene, the two Dirac points are situated at particular points ($\K,-\K$) of the Brillouin zone (BZ), it is important to realize that in general a pair (or several pairs) of Dirac points  may exist at any location $(\D, -\D)$ in the BZ, and may even merge and disappear for appropriate change of band parameters.
Moreover the exceptional  physical properties of graphene-like systems are not only due to the linear dispersion relation but also to the two-component structure of the wave functions corresponding to the two atoms per unit cell. More precisely, each Dirac point may be characterized by a topological number (a Berry phase) which may take two values $\pm \pi$ \cite{Revs}.

\begin{figure}[h!]
\begin{center}
\includegraphics[width=7cm]{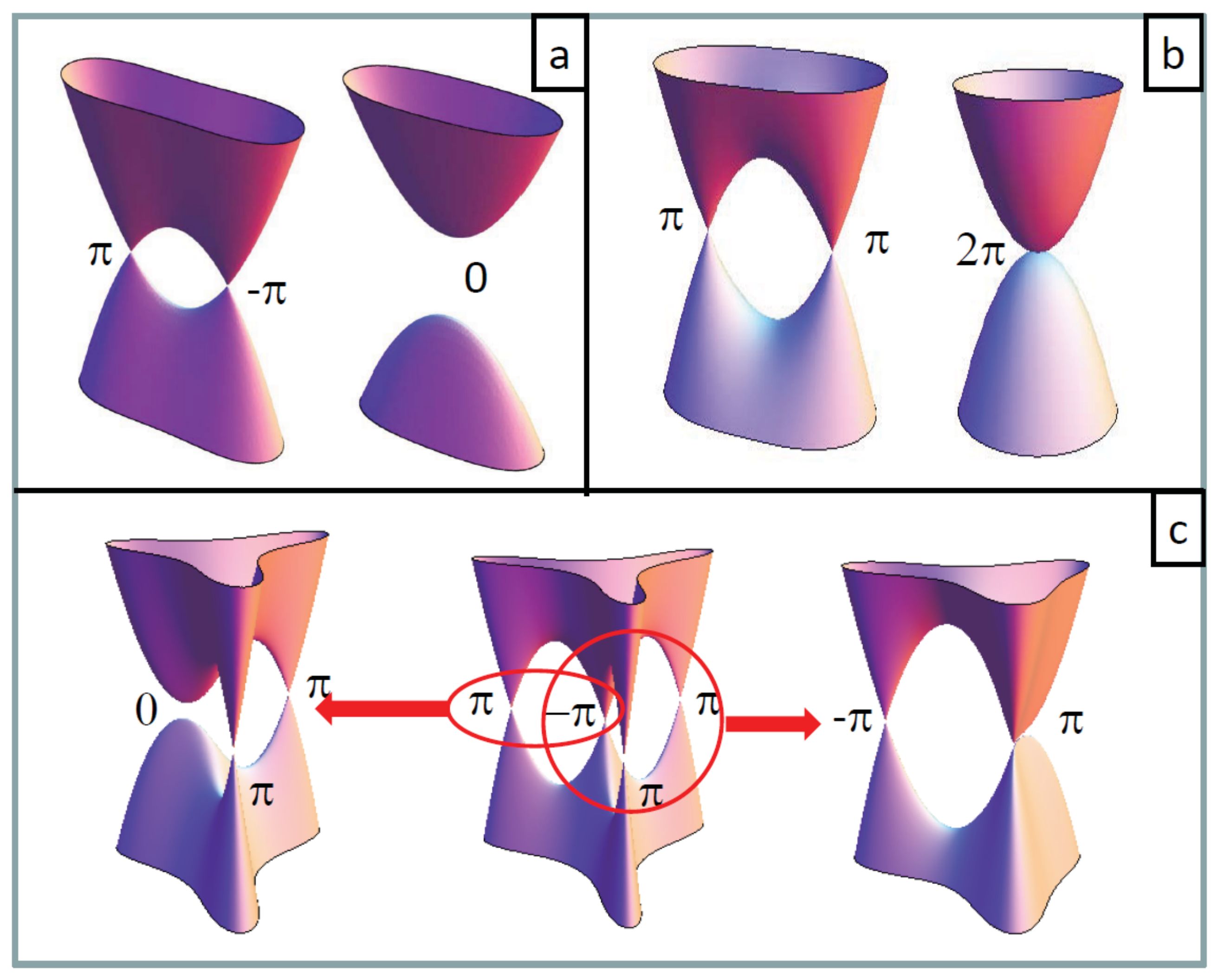}
\caption{Different scenarios for the merging of Dirac cones, corresponding to different topological characteristics.}
\label{fig:mergings}
\end{center}
\end{figure}

In this note, we review different low-energy models that exhibit several Dirac points and discuss different scenarios for the {\it merging} of these Dirac points, according to the topological numbers attached to each point (Fig. \ref{fig:mergings}): a) merging of two Dirac points with opposite Berry phases,  b)  merging of two Dirac points with the same Berry phase, c) a more complex   spectrum which may exhibit different types of merging, in particular a situation that may be called "triple merging" where three Dirac points merge into a single one.

We consider  quite a general structure of a two-component Hamiltonian for a 2D crystal with two  atoms, molecules or orbitals per unit cell. If inversion symmetry is preserved, it has the general form

\begin{equation} {\mathcal H}_\k= \left(
\begin{array}{cc}
                    0 & f_\k \\
                  f^*_\k & 0 \\
                \end{array}
              \right)  \label{GH}  \end{equation}
apart from a term $f_{\k}^0\bone$ that does not affect the spinorial form of the wave functions and that we omit from now on.
The eigenenergies are $\pm |f_\k|$ and the corresponding wave functions are given by
\be \psi = {1 \over \sqrt{2} } \left(
                                 \begin{array}{c}
                                     e^{ i \phi_\k}   \\
                                  \pm 1 \\
                                 \end{array}
                               \right) \ee
where $\phi_\k=\mbox{arg}(f_\k)$. The relative phase $\phi_\k$ exhibits  a particular topological structure: around each Dirac point, the circulation of  $\phi_\k$ along a closed path is quantized such that the  topological Berry phase \cite{Berryphase} defined as
$\phi_B= {1 \over 2} \oint  \nabla_\k \phi_\k \cdot d\k $ takes multiple values of $\pi$. In graphene, the  pair of Dirac points has two      opposite Berry phases $\pm \pi$ . In bilayer graphene, the two Dirac points have a quadratic dispersion relation and the  topological Berry phase is $\pm 2 \pi$ \cite{Berryphase,MF06}. Under appropriate conditions that we discuss later,  each   point  $\pm 2 \pi$ may be separated into two points with a linear dispersion and the  {\it same} Berry phase $(\pi,\pi)$ or $(-\pi,-\pi)$, or even four points with phases
$(\pi,\pi,\pi,-\pi)$ or $(-\pi,-\pi,-\pi,\pi)$ \cite{MF06}.

For the three scenarios mentioned above, we consider the spectrum without and with a magnetic field  and show how it is constrained by topological considerations. In particular, while the {\it maximum} number of  zero-energy Landau levels is given by $w_t=\sum_i |w_i|$,  the number of such levels which are {\it topologically protected} is given by  $w_p=|\sum_i w_i|$,  which plays the role of a minimal number of zero-energy Landau levels. The quantity $w_i= {1 \over 2 \pi} \oint_{C_i}  \nabla_\k \phi_\k \cdot d\k$ is the topological winding number associated with each Dirac point.  Moreover we find that the Landau level spectrum $\{ \ep_n \}$ can be accurately described  by the semiclassical quantization rule

                               \be {\mathcal A}_{\mathcal C}(\ep_n) =2 \pi {e B \over \hbar} ( n + {1 \over 2} - {|w_{\mathcal C}| \over 2} ) \ ,  \label{SCQR}  \ee  where ${\mathcal A}_{\mathcal C}(\ep_n)$ is the area delimited by a closed  contour $\mathcal{C}$ in reciprocal space, associated with energy $\ep_n$.
                               $w_{\mathcal C}$ represents the total winding of the relative phase along this closed contour.

\section{$(\pi, - \pi) \rightarrow 0$ merging}

\begin{figure}[h!]
\begin{center}
\includegraphics[width=8cm]{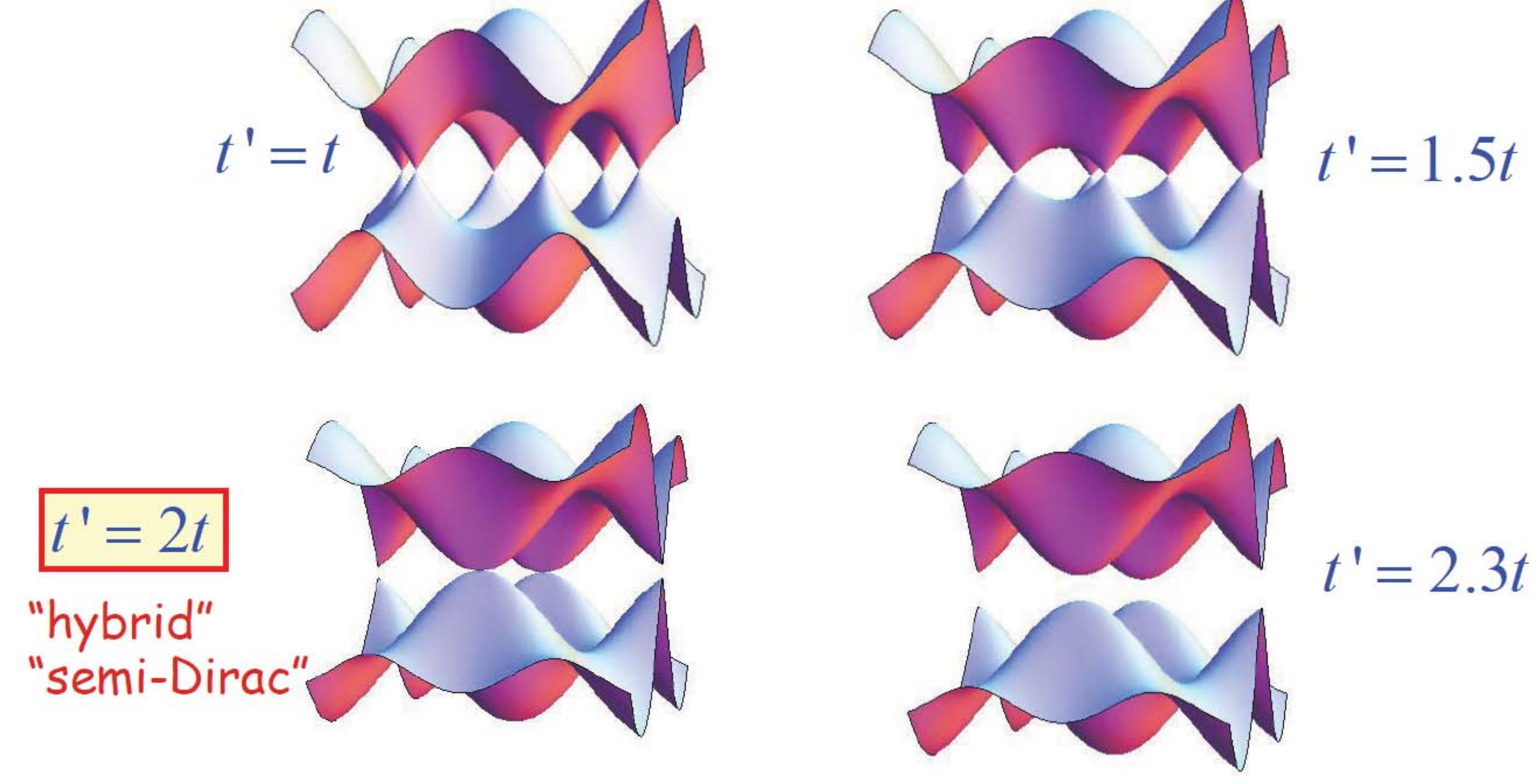}
\caption{Merging of Dirac points in the toy-model of deformed graphene}
\label{fig:fusion-graphenelike}
\end{center}
\end{figure}

We first consider the merging of two massless Dirac points with opposite Berry phases. The simplest model, which exhibits such a transition, is a slight modification of the graphene model, in which one of the transfer integrals is increased \cite{Hasegawa,Dietl,wunsch}:

\be f_\k=t'+ t e^{i \k \cdot \a_1} + t e^{i \k \cdot \a_2}  \ , \ee
  where $\a_1=\frac{a}{2}(\e_x + \sqrt{3}\e_y)$ and $\a_2=\frac{a}{2}(\e_x - \sqrt{3}\e_y)$ are basis vectors of the triangular Bravais lattice,
in terms of the lattice constant $a$.
Figure \ref{fig:fusion-graphenelike} shows the evolution of the spectrum when varying the parameter $t'$. When $t'$ increases, the two Dirac cones approach each other and become anisotropic, until they merge at the time-reversal invariant
$\M$ point when $t'=2t$. At this critical point, the spectrum has a very peculiar structure that we discuss below.

\begin{figure}[h!]
\begin{center}
\includegraphics[width=5cm]{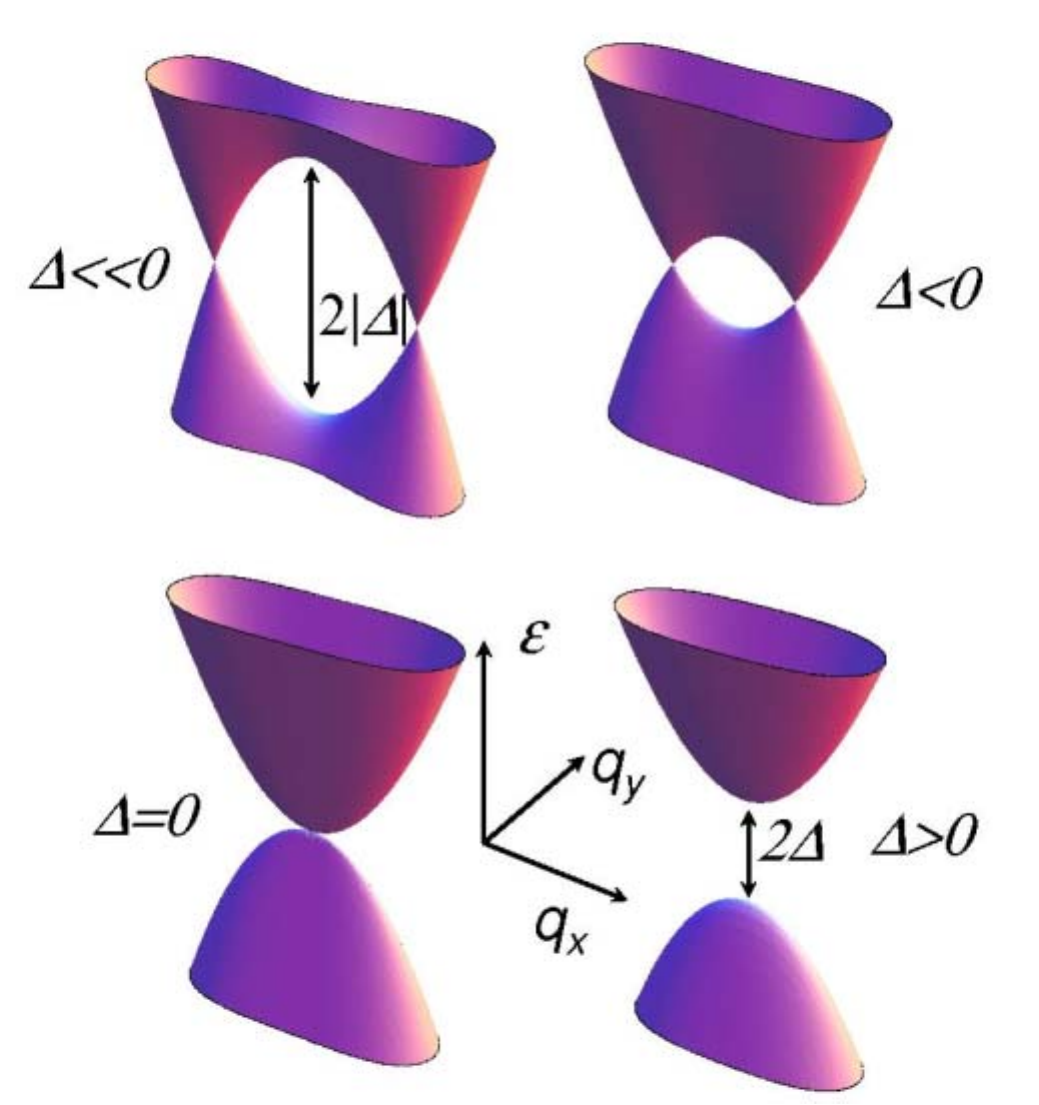}
\caption{Merging of Dirac points described by Hamiltonian (\ref{HU})}
\label{fig:merging-fig1}
\end{center}
\end{figure}

Starting from this toy model we have found more generally under which   conditions, a pair of Dirac points may exist and merge for a   general dispersion               $f_\k= \sum_{mn} t_{mn} e^{i \k . \R_{mn}} $, where the $t_{mn}$ are hopping parameters.
 When time-reversal symmetry is respected, Dirac points come always by pair ($\D,-\D$) and they necessarily merge at a time-reversal invariant point $\G/2$ of the BZ, where $\G$ is a reciprocal lattice vector. In two dimensions, there are  four different possibilities $\G/2= (p \a_1^* + q \a_2^*)/2$, where $p=\pm 1, q= \pm 1$ and $\a_i^*$ are elementary reciprocal lattice vectors. Around these points the Hamiltonian has the form   ($\hbar=1$):

\be {\mathcal H}_\q= \left(
                \begin{array}{cc}
                  0 & -i c q_y + {q_x^2 \over 2 m^*} + \Delta_* \\
                  i c q_y + {q_x^2 \over 2 m^*} + \Delta_* & 0 \\
                \end{array}
              \right) \label{HU} \ee
              where the parameters $m^*, c, \Delta_*$ can be related to the original band parameters \cite{MontambauxUH}.
For example the driving parameter of the transition $\Delta_*$ is given by $\Delta_*= \sum_{m,n} t_{mn} (-1)^{pm+qn}$, where $(p,q)$ are the coordinates of the merging point. Although Hamiltonian (\ref{HU}) has been derived in the vicinity of time-reversal invariant points, it is universal for any merging of type $(\pi,-\pi)$.
When varying the parameter $\Delta_*$, this Hamiltonian describes a topological transition between a semi-metallic phase with two Dirac points and an insulating phase with a gapped spectrum (Fig. \ref{fig:merging-fig1}). At the transition ($\Delta_*=0)$, the spectrum has a ``hybrid" behavior, also called ``semi-Dirac": it is linear in one direction and quadratic in the other
\cite{Dietl,MontambauxUH,Pickett}:

\be \ep_\q= \pm \sqrt{ \left( {q_x^2 \over 2 m_* }\right)^2 + c^2 q_y^2 }  \ . \ee

\begin{figure}[h!]
\begin{center}
\includegraphics[width=7cm]{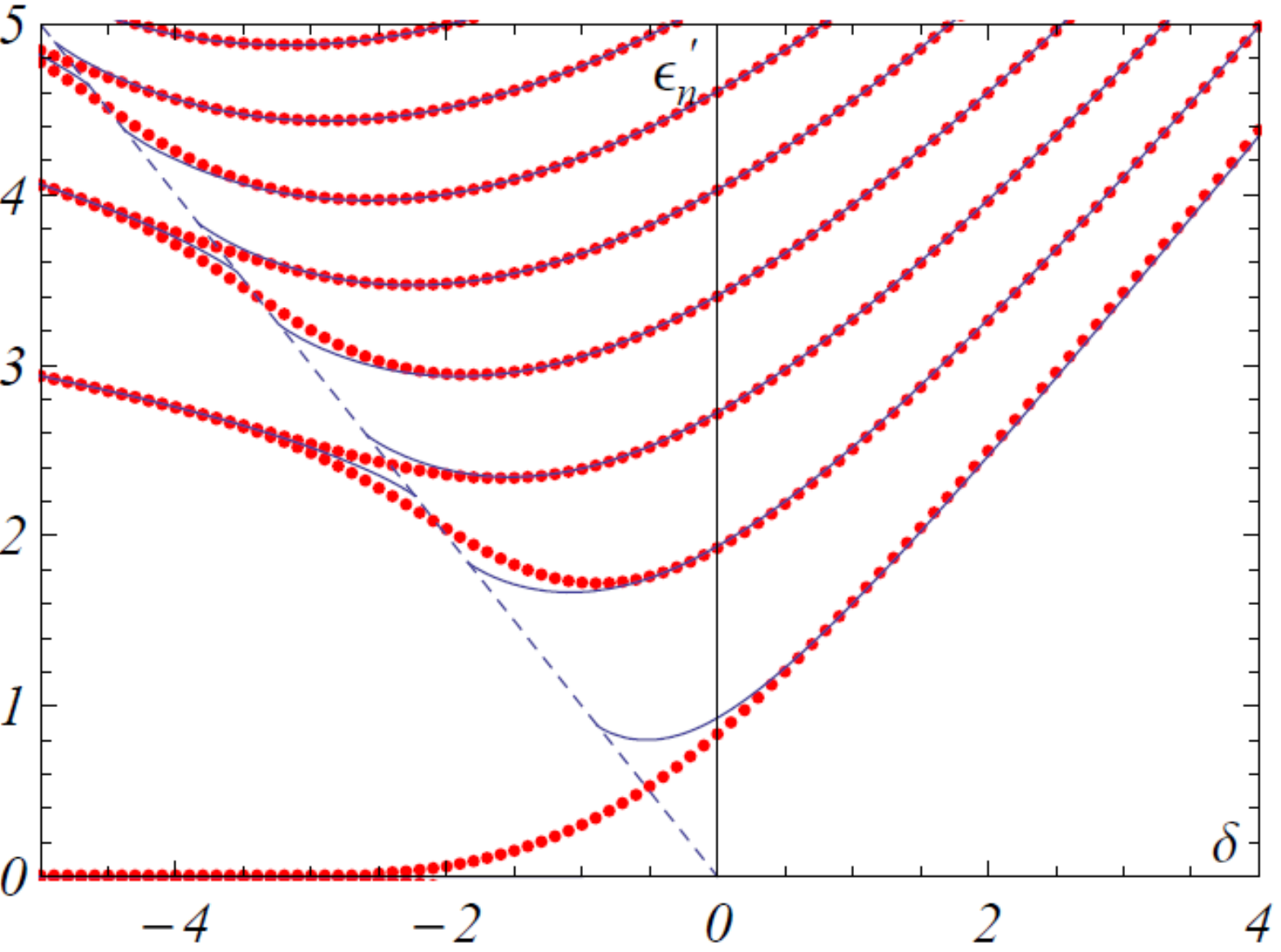}
\caption{Dimensionless energy levels $\ep_n'= \ep_n  \left( {2 m^* \over e^2 \hbar^2 B^2 c^2}\right)^{1/3}$ as a function of the dimensionless parameter $\delta = \Delta_*  \left( {2 m^* \over e^2 \hbar^2 B^2 c^2}\right)^{1/3} \propto \Delta_*/B^{2/3}$.  The blue curves are the results from semiclassical quantization rule (\ref{SCQR}) which is excellent except near the saddle point energy which corresponds to the dashed blue line \cite{MontambauxUH}. Here we have only shown the positive energy levels. The two-fold degeneracy of the zero energy Landau level is lifted when approaching the merging transition.}
\label{fig:merging-fig3}
\end{center}
\end{figure}

 We have studied in detail the Landau level spectrum  in the presence of a magnetic field (Fig. \ref{fig:merging-fig3}). When $\Delta_* \ll 0$, the two Dirac cones are well separated and the Landau spectrum  scales as $\ep_n \propto \pm \sqrt{n B} $ with a two-fold degeneracy corresponding to the two cones with winding numbers $w_{\mathcal C}=\pm1$. When $\Delta_*$ increases, the two valleys start to communicate and the two-fold degeneracy is progressively lifted. At the transition, the spectrum scales as $\ep_n \propto \pm [(n+1/2) B]^{2/3}$, as can be straightforwardly found from the semiclassical quantization rule (\ref{SCQR}), where the winding number $|w_{\mathcal C}| = 0$ since the two Dirac points have merged and their opposite Berry phases are annihilated. Moreover the spectrum is clearly separated in two different regions by the saddle point connecting the two Dirac cones. Below the saddle point a closed  contour at a given energy encloses only one Dirac point and $w_{\mathcal C}=\pm 1$. Above the saddle point energy, such a contour encloses the two Dirac points and $w_{\mathcal C}=0$. As shown in Fig. \ref{fig:merging-fig3}, the semiclassical quantization rule (\ref{SCQR}) is excellent, except for the vicinity of the saddle point because of the discontinuous change in the area ${\mathcal A}_{\mathcal C}$ and the
associated change in the total winding number $w_{\mathcal C}$ in Eq. (\ref{SCQR}).

In graphene, such a merging of Dirac points would require a modification of hopping integrals impossible to reach experimentally \cite{strain}.
However, Dirac cones exist in other physical systems, such as organic 2D conductors \cite{organics}, optical lattices of cold atoms \cite{cold-atoms}, where the merging transition may be more easily achieved.

\section{$(\pi,  \pi) \rightarrow 2 \pi$ merging}

We now consider the case of  bilayer graphene in which the two layers   have a  translational or   rotational mismatch (twist) with respect to the perfectly A-B stacked case  \cite{tilted-bilayer}. In a first approximation (which will be discussed in the next paragraph), the low energy Hamiltonian  around the $\K$ point  can be written in the form \cite{deGail111}

\be \mathcal{H}_\q= \left(
                \begin{array}{cc}
                  0 &  {\Pi^2 \over 2 m^*} - \Delta \\
                 {{\Pi^*}^2 \over 2 m^*} - \Delta & 0 \\
                \end{array}
              \right) \label{HBL} \ee
where $\Pi = q_x+i q_y$. When $\Delta=0$, the spectrum is parabolic and the associated Berry phase is $2 \pi$  ($-2 \pi$ around the $-\K$ point). For non-zero values of $\Delta$, which we have chosen to be real,
the quadratic Dirac cone splits into two linear Dirac cones,   though with the {\it same} Berry phase $\pi$,
in contrast to the $(\pi,-\pi)$ universality class where the relative sign in the Berry phases was dictated by time-reversal symmetry.
Two fundamental differences with the previous case are noticeable. First the zero field spectrum stays gapless for all  $\Delta$, \cite{signDelta} second the spectrum in a magnetic field exhibits $w_p=2$   modes of zero energy ( $\times 4$ to account for valley and spin degeneracy) which, in contrast to the previous merging, are topologically stable.

\begin{figure}[h!]
\begin{center}
\includegraphics[width=7cm]{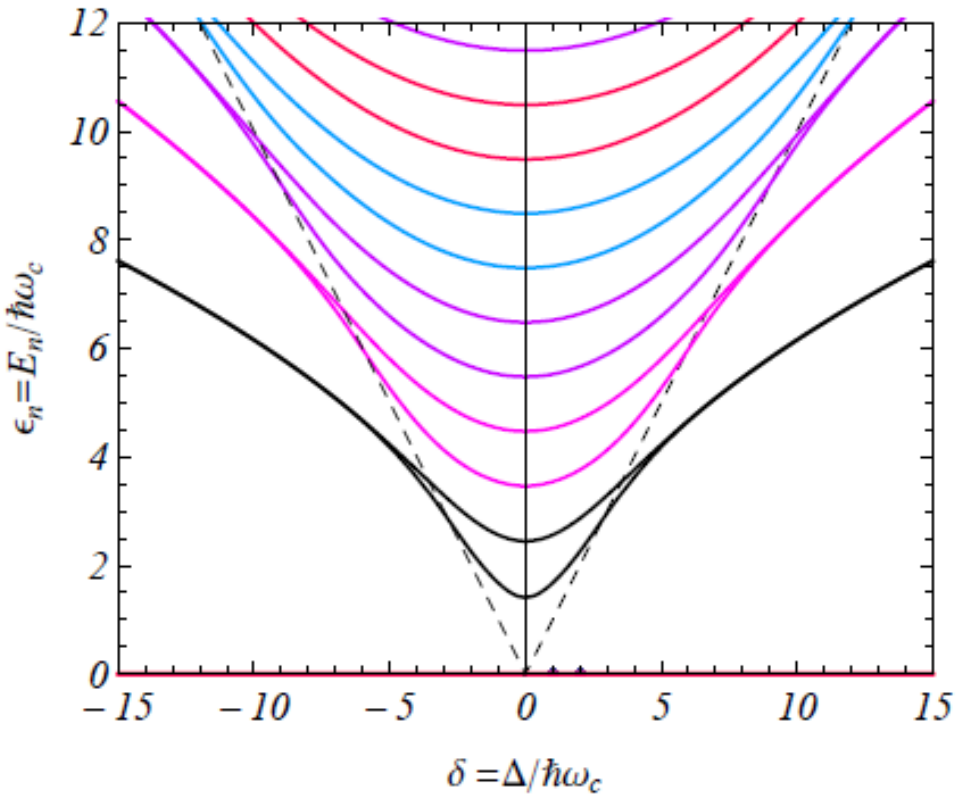}
\caption{Landau levels in the model (\ref{HBL}) of bilayer graphene $((\pi,\pi)\rightarrow 2 \pi$ merging). The dashed lines indicate the energy of the saddle point, for more details see \cite{deGail111}. This spectrum has to be compared with the spectrum  corresponding to the $(\pi, -\pi)\rightarrow 0$ merging (Fig.~\ref{fig:merging-fig3}). Here, the zero energy levels stay topologically stable.}
\label{fig:landau-bilayer-tilted}
\end{center}
\end{figure}

The Landau level spectrum  have been calculated in \cite{deGail111}. In Fig. \ref{fig:landau-bilayer-tilted}, it is shown as a function of the dimensionless parameter $\Delta/\hbar \omega_c$ with $\omega_c= e B / m^*$. At large $\Delta$, the spectrum scales as $\pm \sqrt{n B}$, as expected from the semiclassical quantization rule with $w_{\mathcal C}=1$. When $\Delta$ is small, or well above the saddle point energy,  one expects to recover the spectrum of the A-B stacked bilayer, $\pm \hbar \omega_c \sqrt{n(n-1)}$, which in the limit of large $n$ scales as $\pm \hbar \omega_c (n-1/2)$ as expected from Eq.~(\ref{SCQR}) with $w_{\mathcal C}=2$.
  Notice that, in contrast to monolayer graphene where the saddle points occur at a rather high energy  ($\sim 3$ eV),
the energy of the saddle points in twisted bilayer graphene depends on the rotation angle between the layers and may be found in
the $10-100$ meV regime \cite{Eva}.

\section{Multimerging}

It turns out that the spectrum in bilayer graphene has an interesting fine structure at low energies. Due to   interlayer hopping between atoms which do not face each other, the low-energy Hamiltonian takes indeed the form

\be \mathcal{H}_\q= \left(
                \begin{array}{cc}
                  0 &  {\Pi^2 \over 2 m^*}+ v \Pi^* - \Delta \\
                 {{\Pi^*}^2 \over 2 m^*}+ v \Pi - \Delta & 0 \\
                \end{array}
              \right) \label{Htri} \ee
where $v$ is a small perturbation  (neglected in the previous section) responsible for the trigonal warping of the isoenergy contours at low energies \cite{MF06}. The
parameter $\Delta$, which we choose to be real here, mimics a strain of the bilayer or a displacement between the layers \cite{coreans},
with respect to perfect AB stacking ($\Delta=0$).
               \begin{figure}[h!]
              \begin{center}
\includegraphics[width=6cm]{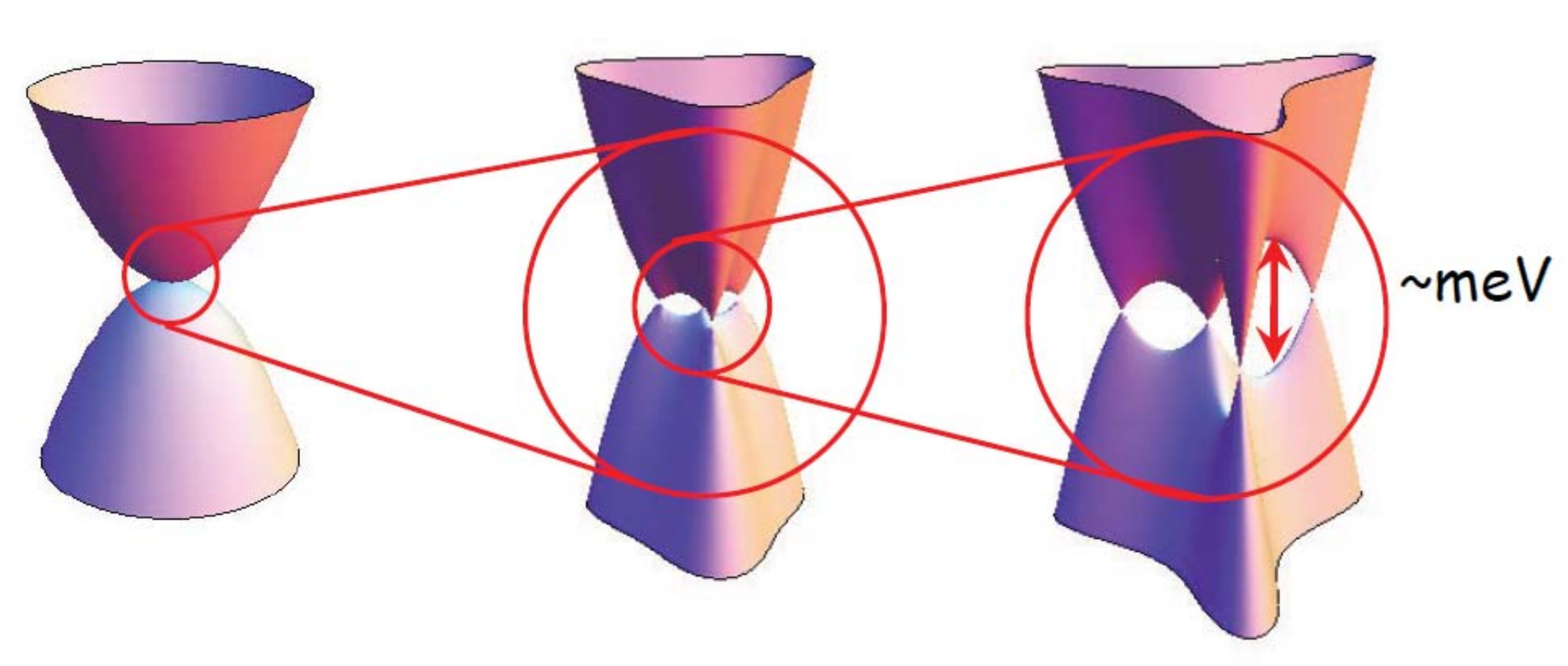}
\caption{At low energy, the spectrum of graphene bilayer is not exactly quadratic, it has a fine structure made of four Dirac points around the each of the two points $(\K, - \K)$ of the BZ. }
\label{fig:zoomintrigonal}
\end{center}
\end{figure}
Fig. \ref{fig:zoomintrigonal} shows the energy spectrum when $\Delta=0$. It consists of a central massless Dirac point surrounded with three massless Dirac points of opposite winding number, separated by (three) saddle points at energy $m_* v^2/2$. A non-zero value of $\Delta$ may induce topological transitions of two kinds (Fig. \ref{fig:mergings}-c). When $\Delta <0$, the central Dirac point merges with one of the three other points, leaving the other two unaffected. This transition is of the first kind mentioned above ($(-\pi, \pi) \rightarrow 0$) and is properly described by Hamiltonian (\ref{HU}). When $\Delta >0$ , the central Dirac point merges with two other points, leaving the fourth one unaltered \cite{coreans}. Fig. \ref{fig:trimerging1} shows the evolution of the relative phase $\phi_\k$ and of the Berry phases attached to each point, in the two situations.

\begin{figure}[h!]
\begin{center}
\includegraphics[width=7cm]{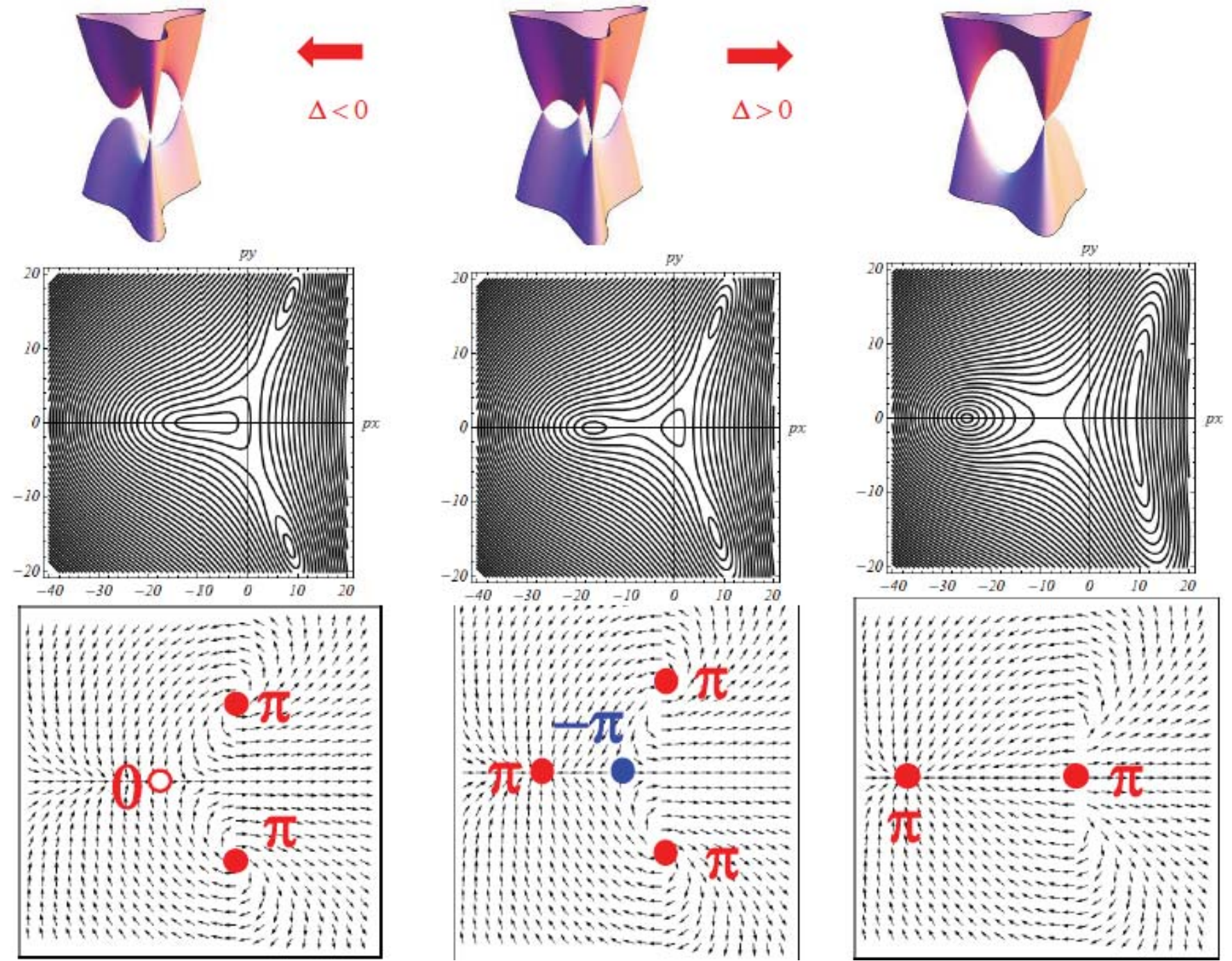}
\caption{Dependence of the phase $\phi_\k$ for the Hamiltonian (\ref{Htri}). }
\label{fig:trimerging1}
\end{center}
\end{figure}

\begin{figure}[h!]
\begin{center}
\includegraphics[width=8cm]{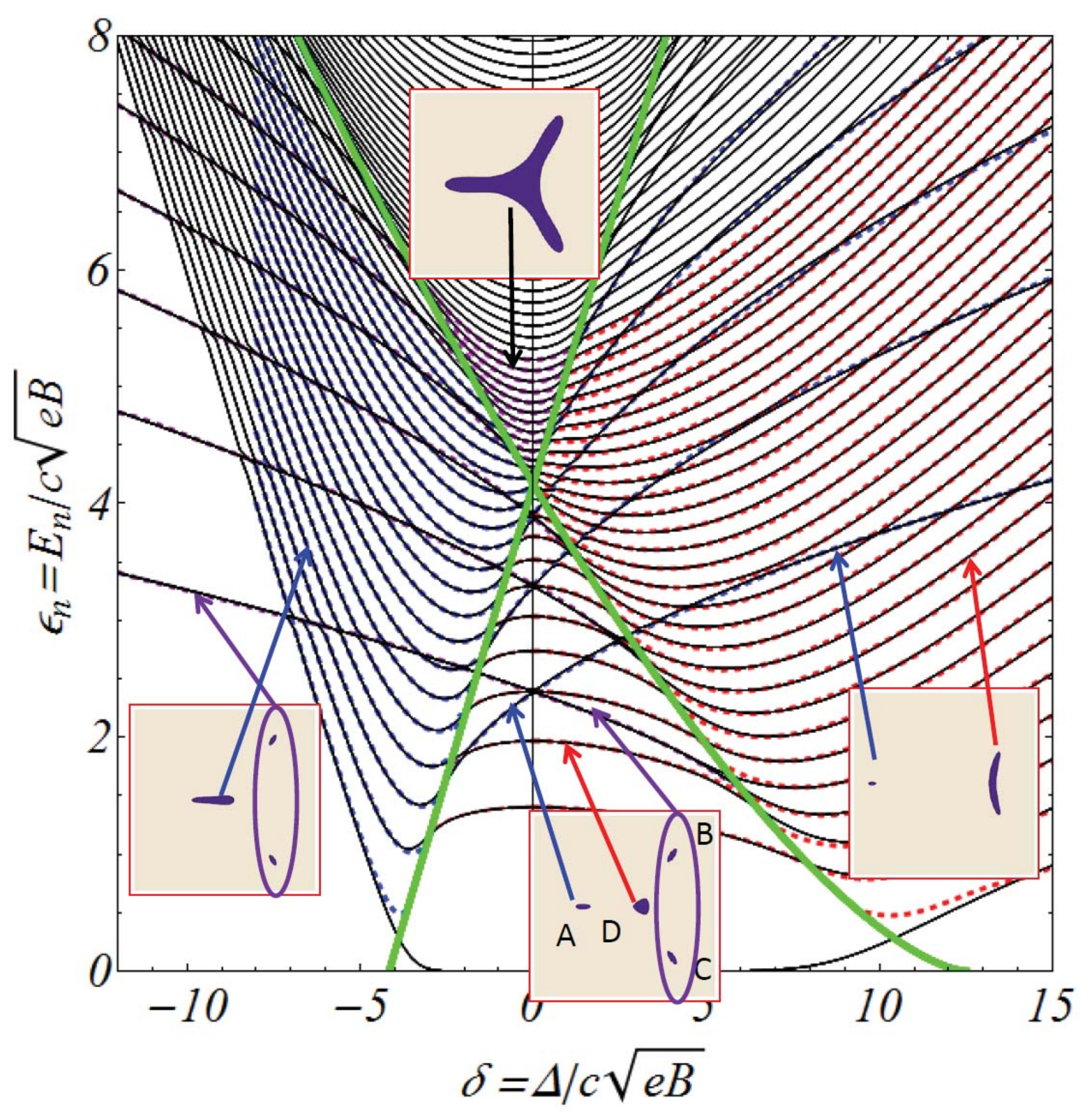}
\caption{Landau level spectrum for the Hamiltonian (\ref{Htri}) as a function of the parameter $\Delta$. In the central low energy spectrum, there are $w_t=4$ zero energy modes corresponding to the four distinct massless Dirac cones. When $\Delta$ increases, two of these modes get a finite (positive and negative) energy, leaving only $w_p=2$ topologically protected modes.}
\label{fig:landau-multimerging}
\end{center}
\end{figure}

\begin{figure}[h!]
\begin{center}
\includegraphics[width=8cm]{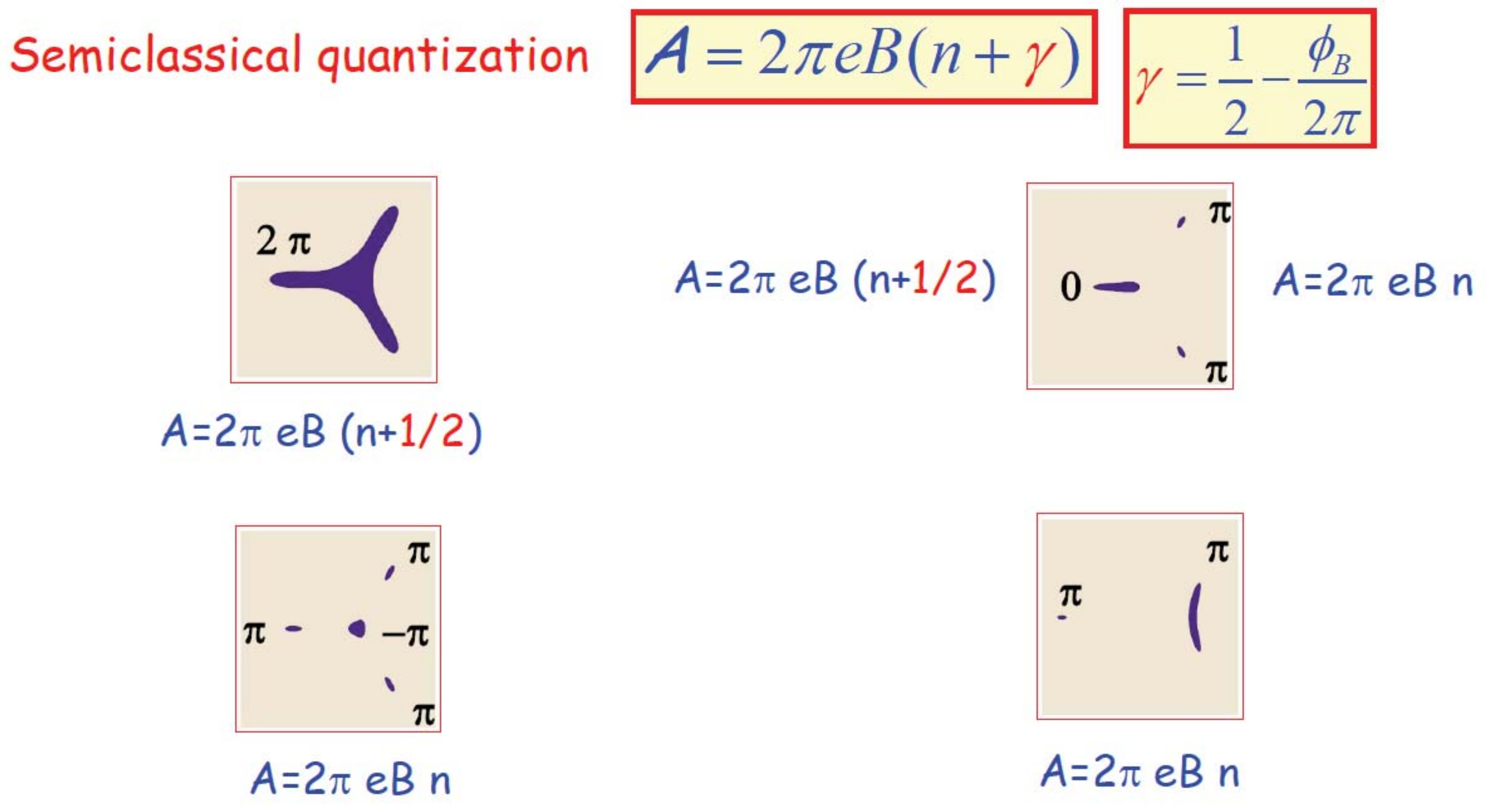}
\caption{Semiclassical quantization of the different pockets involved, with different Berry phases, eq. \ref{SCQR}.}
\label{fig:quantization-pockets}
\end{center}
\end{figure}

 The Landau level spectrum  in this model has been calculated recently \cite{Falko11}. We have given a full description of this very rich spectrum, together with a semiclassical analysis \cite{deGail112}. Figure \ref{fig:landau-multimerging} shows a comparison between the numerically obtained spectrum (full lines)
and that (dashed lines) derived from the semiclassical quantization rule (\ref{SCQR}) as a function of the mismatch parameter $\Delta$.
The spectrum can be divided into four distinct regions,  according to the connectivity of the Fermi surface, separated by the two green curves which represent the energies of the saddle points connecting adjacent Dirac cones. The insets represent the pockets delimited by a constant energy in the different regions of the spectrum. By semiclassical quantization (\ref{SCQR}) of these pockets, we have been able to fit perfectly (except near the saddle points, where semiclassical quantization breaks down) the  full Landaulevel spectrum. The different sets of levels are related to the different pockets. Applying the quantization rule (\ref{SCQR}), it is essential to attribute the correct winding number $w_{\mathcal C}$ (or Berry phase) to the different pockets. This is illustrated in  Fig.~\ref{fig:quantization-pockets}.
Finally, we find that in the central low-energy region, there are four zero-energy Landau levels (for each valley and each spin direction) corresponding to the four Dirac points. Among these four modes, only two are topologically protected, corresponding to the total algebraic Berry phase $w_p=2$ attached to these four points.

\section{Zero modes and semiclassical quantization rule}

In conclusion, we provide a heuristic argument relating the total number of topologically protected zero-energy modes to the semiclassical quantization rule.
Consider first the   Hamiltonian  \cite{Volovik,Manes}
\be {\mathcal H}= \lambda \left(
                            \begin{array}{cc}
                              0 & \Pi^{p} \\
                              {\Pi^*}^{p} & 0 \\
                            \end{array}
                          \right)
                          \ee
describing a Dirac point with energy spectrum $\ep =\pm \, \lambda |\q|^{p}$ and with a winding number $p$ (a topological Berry phase $\phi_B= \pm p \pi$). In a magnetic field, performing the substitution $\Pi \rightarrow \sqrt{2 e B}\, a$ where $a$ is the annihilation operator, it is straightforward to find that the Landau level spectrum in a magnetic field is given by $\ep_n(B)= \pm \lambda (2 e B)^{p/2} \sqrt{n(n-1) \cdots (n-p+1)}$. In the limit of large $n$, this spectrum approximates as $\ep_n(B)\simeq \pm \lambda [2 e B(n+{1 \over 2} - {p \over 2})] ^{p/2} $. This corresponds to the semiclassical quantization rule (\ref{SCQR}), $w_{\mathcal C}=w_t=w_p$ being equal to the total winding number $p$.

Consider now a Hamiltonian describing $p$ massless Dirac points with Berry phase $\pi$ and $p'$ massless Dirac points with Berry phase $-\pi$.  It can be written as (\ref{GH}) with $
f_\q= \lambda \,   \prod_{j=1}^{p'} (\Pi^* - \beta_j^*) \prod_{i=1}^{p} (\Pi - \alpha_i)$
,
where $\alpha_i$ and $\beta_j$ are the complex positions in reciprocal space of the $\pi$-phase and $-\pi$-phase Dirac points,
respectively.
The total number of Dirac points, and thus the maximal number of zero-energy Landau levels,  is $w_t= p+p'$. In order to find the total number of topologically protected zero-energy levels, we continuously modify the parameters  $\alpha_i, \beta_j\rightarrow 0$ so that $f_\k$ becomes $ {\Pi^*}^{p'} \Pi^{p}$.   In a magnetic field, assuming for example that $p >p'$, this term keeps the form $\sqrt{2 e B}^{w_t} (a^\dagger a)^{p'} a^{p-p'}$ indicating that there are $w_p=p-p'$ zero energy levels. Moreover in this case, it is straightforward to show that the energy levels scale as $\sqrt{n(n-1) \cdots (n- p+p'-1)}$ so that the large-$n$ expansion reproduces the semiclassical quantization rule with $w_p= p-p'$.  Therefore although the maximum number of zero-energy Landau levels is
$w_t=p+p'$, a quantum-mechanical coupling between them partially lifts the degeneracy, but $w_p=|p-p'|$ zero modes remain
topologically protected.

\bigskip

This work was partially supported by the
NANOSIM-GRAPHENE project (ANR-09-NANO-016-01)
of ANR/P3N2009.

% \appendix

% \section{}
% \label{}


\begin{thebibliography}{00}
\bibitem{Berryphase}
 The quantity $\phi_B$ defined here is not exactly the real  Berry phase and is named as the topological Berry phase. It is actually a winding number ($\times \pi$) . The real Berry phase is only meaningfull modulo $2\pi$, see Refs.~\cite{Fuchs,Park11}. Here, for simplicity, we keep the name Berry phase for this quantity.
\end{thebibliography}
\end{document}